\documentstyle[11pt,aaspp4]{article}

\lefthead{Shabanova et al.}
\righthead{Free precession in PSR B1642$-$03}

\begin{document}
\title{Evidence for Free Precession in the Pulsar B1642$-$03}

\author{T.V. Shabanova\altaffilmark{1}}
\affil{Astro Space Center, P.N.Lebedev Physical Institute, 53 Leninskij
       Prospect, Moscow 117924, Russia}

\author{A.G. Lyne\altaffilmark{2}}
\affil{University of Manchester,
       Jodrell Bank Observatory, Macclesfield, Cheshire SK11 9DL, UK}

\and

\author{J.O. Urama\altaffilmark{3}}
\affil{Hartebeesthoek Radio Astronomy Observatory (HartRAO),
       P.O. Box 443, Krugersdorp 1740, South Africa}

\affil{Department of Physics \& Astronomy, University of Nigeria,
       Nsukka, Enugu State, Nigeria}

\altaffiltext{1}{e-mail: tvsh@prao.psn.ru}
\altaffiltext{2}{e-mail: agl@jb.man.ac.uk}
\altaffiltext{3}{e-mail: johnson@hartrao.ac.za}

\begin{abstract}
We present an analysis of the timing data of the pulsar
B1642$-$03, collected over a span of 30 years between 1969 and
1999. During this interval, the timing residuals exhibit cyclical
changes with amplitude varying from 15 ms to 80 ms and spacing of maxima
varying from 3 years to 7 years. Interpretation of these observed
cyclical changes in terms of free precession suggests a wobble
angle of about 0.8 degrees.
\end{abstract}


\section{Introduction}
The pulsar B1642$-$03 was discovered more than 30 years ago
(\cite{hug69}) and is known to exhibit interesting timing
behaviour. It has a period of 0.387 s and a spindown rate
$1.78\times10^{-15}$ s/s, indicating that it is a relatively young
pulsar, with a characteristic age of ${P}/2{\dot{P}} \sim 3.4$ Myr.

Since the pulsar discovery, timing measurements were carried out
almost continuously for 13 years between 1969 July and 1982
September at the frequency of 2388 MHz using antennas of the Deep
Space Network of NASA (\cite{dow83}; \cite{dow86}). From a timing
analysis of the JPL data, Cordes and Downs (1985) found that the
timing residuals of PSR B1642$-$03 show oscillatory behaviour with
an amplitude of about 15 ms. Subsequently it was revealed that PSR
B1642$-$03 also exhibits periodic changes in the shape of the mean
pulse profile with a period of about 1000 days (\cite{bla91}).
According to Cordes (1993), cyclical timing residuals and periodic
changes in the shape can be explained by the precession model with
a wobble angle ${\theta_p}{\sim}0.5\arcdeg$. This interpretation
implies that free precession yields pulse shape changes from wobble
of the beam besides any wobble that is also manifested in timing
residuals because of the spindown torque law.

In this paper we continue to investigate the timing behaviour of
the pulsar B1642$-$03, analyzing the data collected for 30 years.
The extension of the timing data span from 13 years to 30 years
has become possible due to regular pulsar timing observations,
which were carried out at Jodrell Bank Observatory in the UK, at
Hartebeesthoek Radio Astronomy Observatory in South Africa and at
Pushchino Radio Astronomy Observatory in Russia.  The Jodrell Bank
data were obtained in the wide frequency range 0.4 - 1.6 GHz and
cover a span of 18 years between 1981 July and 1999 September. The
HartRAO timing data obtained at 1.6/2.3 GHz cover a span of
$\sim$14 years between 1985 November and 1999 March.  The
Pushchino data were obtained at 0.1 GHz and include the timing data
between 1991 March and 1999 October and a separate observing
session in 1984 September - December.  Together, these provide a
continuous data set collected for 30 years with a one year gap
between 1983 July and 1984 August. A timing analysis of these data
reveals that the timing residuals of PSR B1642$-$03 exhibit
cyclical behaviour over the entire time span of observations.  The
best explanation of the observed cyclical timing residuals is
probably provided by the precession model.  Partly these results
were earlier published in Proc. IAU Col. 177 (\cite{sha00}).

\section{Observations}
At Jodrell Bank, observations of the pulsar with the 76-m Lovell
radiotelescope started in July 1981.  Cryogenic receivers
sensitive to the two hands of circular polarisation were used at
frequencies centred on 408, 610, 1400 and 1600 MHz with observing
bandwidths of 8 MHz at the lower two frequencies and 32 MHz for
the two highest.  Each observation typically consisted of 6
1-minute sub-integrations.  The profiles from these were added in
polarisation pairs and then combined to provide a single
total-intensity profile.  This was then convolved with a template
derived from a single high signal-to-noise ratio profile at the
same frequency to give a time-of-arrival.

At HartRAO, the timing observations of the pulsar B1642$-$03
commenced in 1985 November and were performed with the 26-m
radiotelescope at frequencies near 1.6 or 2.3 GHz, using the
system described by Flanagan (1993). The observing bandwidth was
10 MHz. Each observation usually consisted of three integrations.
Pulse arrival times were obtained from 12 min on-line integrations
of the pulsed signal sampled at 0.15 ms intervals, using a filter
time-constant of 200 microsecs. The pulse was approximated by a
Gaussian. Such Gaussian templates have a reference point, which
was chosen to be the centre of the main component.

The pulsar timing observations at Pushchino Observatory were
started in 1991 March using the BSA radiotelescope, making up a
linearly polarized transit antenna, which operated at 102.7 MHz
until 1998 May and at 111.3 MHz since 1998 November after the BSA
reconstruction. A 32-channel receiver with a channel bandwidth of
20 kHz was used for the observations. The receiver time constant
was 3 ms. Each measurement consisted of 3.6 minute integration of
the pulsar signal, synchronized with the apparent pulsar period
(\cite{sha98}). Observation time corresponded to transit time at
the declination of the pulsar. The data consist of the mean pulse
profiles recorded approximately three times a month. Pulse arrival
times were derived by cross-correlating the mean pulse profile
with a standard, low-noise template.

\begin{figure}
   \plotone{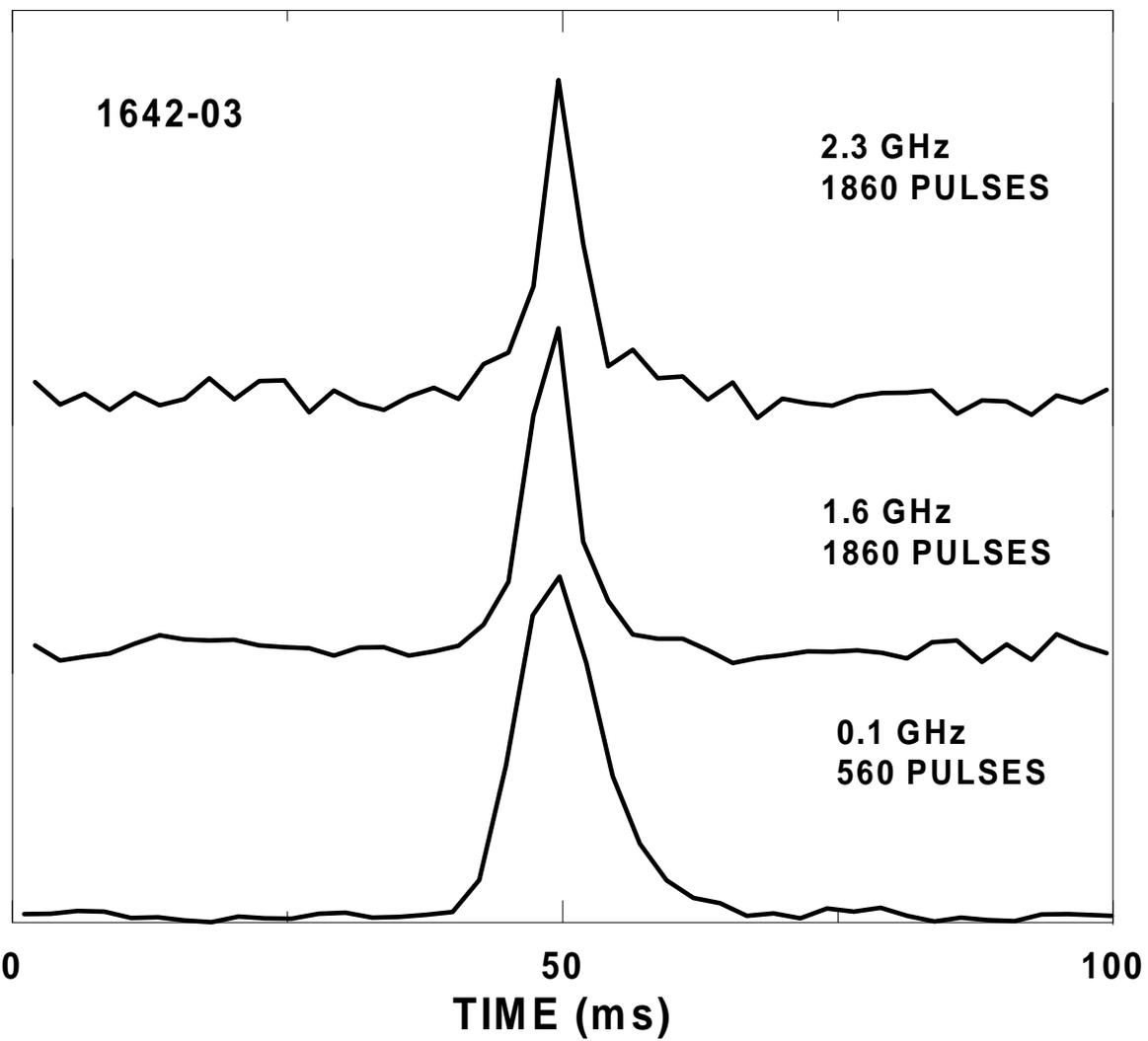}
   \caption{The mean pulse profile of the pulsar B1642$-$03
           for a single observation at different frequencies of 0.1,
           1.6  and 2.3 GHz. The alignment between the profiles is
           arbitrary.
           \label{profil}}
   \end{figure}

The mean pulse profiles of the pulsar B1642$-$03 for a single
observation at different observing frequencies are shown in
Figure~\ref{profil}.

\placefigure{profil}

\section{Timing analysis and results}
The topocentric arrival times collected at JBO, HartRAO and PRAO
and the geocentric arrival times obtained from the archival JPL
timing data (\cite{dow83}; \cite{dow86})  were all referred to the
Solar System barycenter at infinite frequency using a standard
pulsar timing technique (\cite{man77}). This involved the use of
the JPL DE200 ephemeris and the position given by Downs \&
Reichley (1983), together with a proper motion equal to zero. A
second-order polynomial describing secular spin-down pulsar
behaviour was fitted to the barycentric arrival times to obtain
residuals from a timing model. The timing residuals derived as the
observed times minus the predicted ones were used for improving
the astrometric and the spin-down parameters of the pulsar.
Measured timing parameters are shown in Table~\ref{tbl-1}.
The period and the period derivative were determined from
the entire interval of observations from 1969 to 1999
(fit interval MJD 40414 - 51450).
Analysis of the timing residuals showed that the pulsar has a
proper motion close to zero. Recent analysis
of Martin (1999)\nocite{mar99} confirmed that the proper
motion is $12\pm 8$ milliarcsec/year and barely significant.
The pulsar catalog (\cite{tay93}) quotes a proper motion obtained
by Lyne, Anderson, \& Salter (1982)\nocite{lyn82} as being much larger.
However, their measurements have large errors and this, in principle,
does not exclude a possibility that the proper motion may be zero.
The observations of the pulsar in the range  0.1 - 1.6 GHz
permitted to determine the dispersion measure with high accuracy.

\placetable{tbl-1}

\subsection{Timing residuals of PSR B1642$-$03}
\begin{figure}
   \plotone{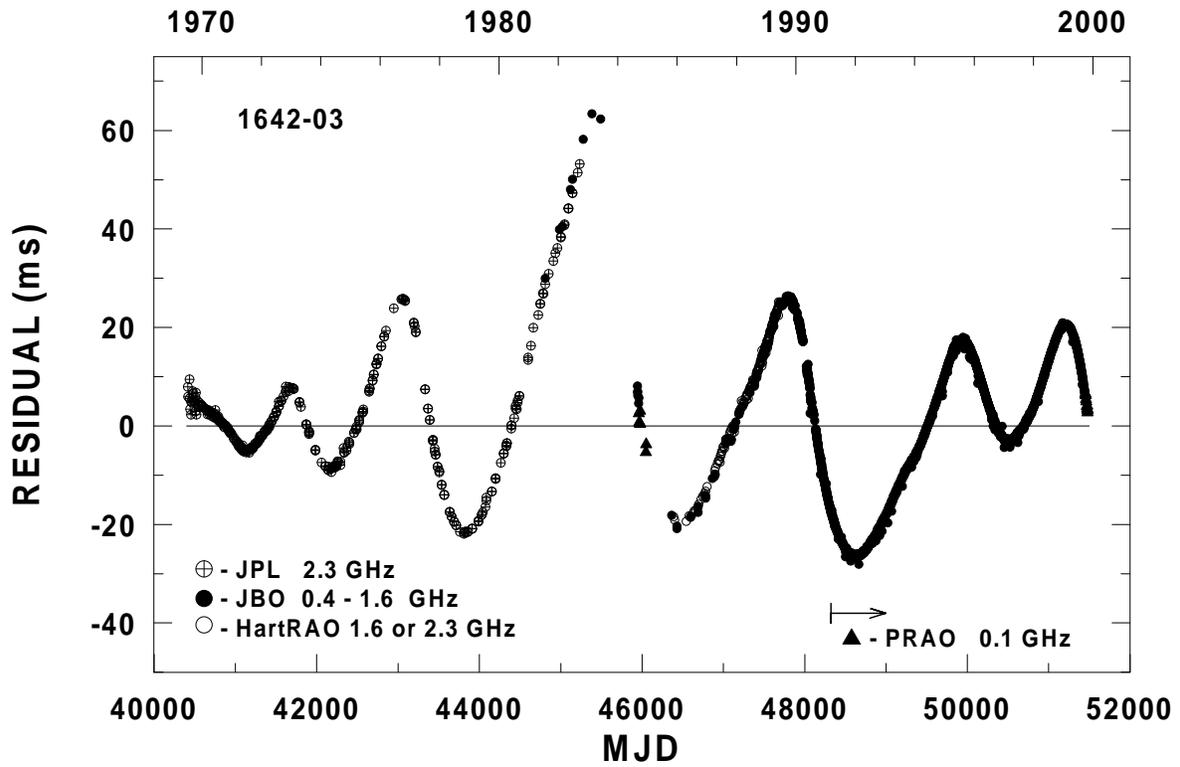}
   \caption{The timing residuals for PSR B1642$-$03 from
           the combined JPL, JBO, HartRAO and PRAO data set over
       the interval from 1969 to 1999.
           \label{phres}}
   \end{figure}

Figure~\ref{phres} shows the timing residuals of PSR B1642$-$03
after removing the second-order polynomial from the combined JPL,
JBO, HartRAO and PRAO data set of the barycentric arrival times.
As can be seen, the timing residuals for PSR B1642$-$03 exhibit
oscillatory behaviour and more than 6 cycles are observed over the
entire 30-yr data span. The central part of the residual curve has
a one year gap between 1983 July and 1984 August, but
nevertheless, it gives adequate information. The curve presents
fluctuations whose amplitudes initially increase and then fade.
The amplitude of the cycles varies between 15 ms and 80 ms and the
spacing of maxima varies between 3 years and 7 years. It will be noticed
that the larger amplitudes correspond to the larger spacings.
The curve is asymmetric with respect to the axis X.

The observed cyclical changes are atypical of most timing noise seen
from radio pulsars. Most timing noise is consistent with red, stochastic
noise that is aperiodic. The timing residuals from PSR B1642$-$03 are not
strictly periodic but are sufficiently cyclical that it is plausible
to interpret them as the result of a wobble of the spin axis of the
pulsar. Figure~\ref{phres} demonstrates that the shape of
the residual curve is robust against the particular length of the data
set that is analyzed. The general character of the variations which
was found earlier by Cordes \& Downs (1985, Fig.10l)\nocite{cor85}
for the interval 1969 - 1981 remained the same in our plot despite
the extension of our data span to 30 years.

\placefigure{phres}

\subsection{Analysis of the spin-down parameters of PSR B1642$-$03}
\begin{figure}
   \plotone{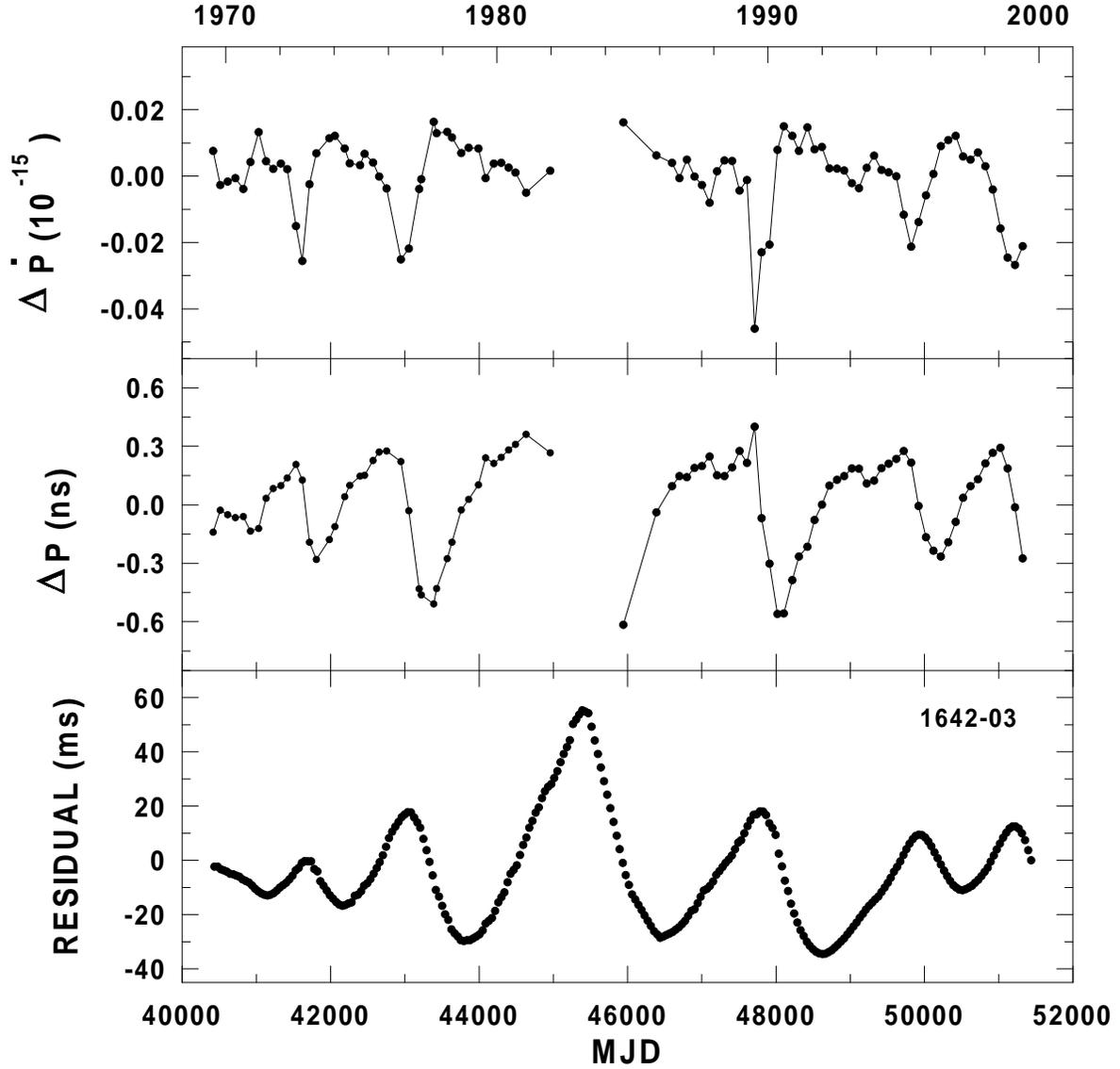}
   \caption{The period derivative residuals,
          $\Delta{\dot{P}}$, the period residuals, $\Delta{P}$, and
          the timing residuals over 30 years of observations relative
          to the spin-down model given in Table~\ref{tbl-1}.
           \label{rppdot}}
   \end{figure}

For a more detailed analysis of the variations, we demonstrate
changes of the spin-down parameters $P$ and $\dot{P}$ versus time
together with the timing residuals in Figure~\ref{rppdot}. The
rotation period, $P$, and the period derivative, $\dot{P}$, were
calculated from the local fits, performed over intervals of 200
days which overlapped by 100 days. The timing residuals shown in
the lower plot were transformed into a uniform curve by averaging
the residuals over 40-day intervals. The empty intervals were filled
with numbers, derived by linear interpolation of average values of
residuals taken from the nearest adjacent intervals. Such a
procedure did not deform the initial curve, since the averaging
time was considerably less than the timescale of cyclic changes.
As is seen, the plotted period residuals, $\Delta{P}$, and the
period derivative residuals, $\Delta{\dot{P}}$, exhibit cyclical
behaviour and the observed variations in the period derivative
$\dot{P}$ make up about 2$\%$ of the mean value of
$1.78\times10^{-15}$. The variations in some cycles are
similar in form, although these cycles do not exhibit a strict
periodicity in their recurrence. Besides, the gap in the centre of
these plots makes the picture of the changes incomplete.
The variations in rotation parameters can cause the
variations in the pulse shape of the pulsar. An analysis of the
JBO data to reveal long-term variations in the pulse shape showed
no significant pulse profile changes. However the precision of the
measurements does not preclude the possibility of changes of the
magnitude claimed by Blaskiewicz (1991)\nocite{bla91}. Any pulse
shape changes at the low frequency of 0.1 GHz may not be noticed
because the mean pulse profile has a simple shape.

\placefigure{rppdot}

\begin{figure}
   \plotone{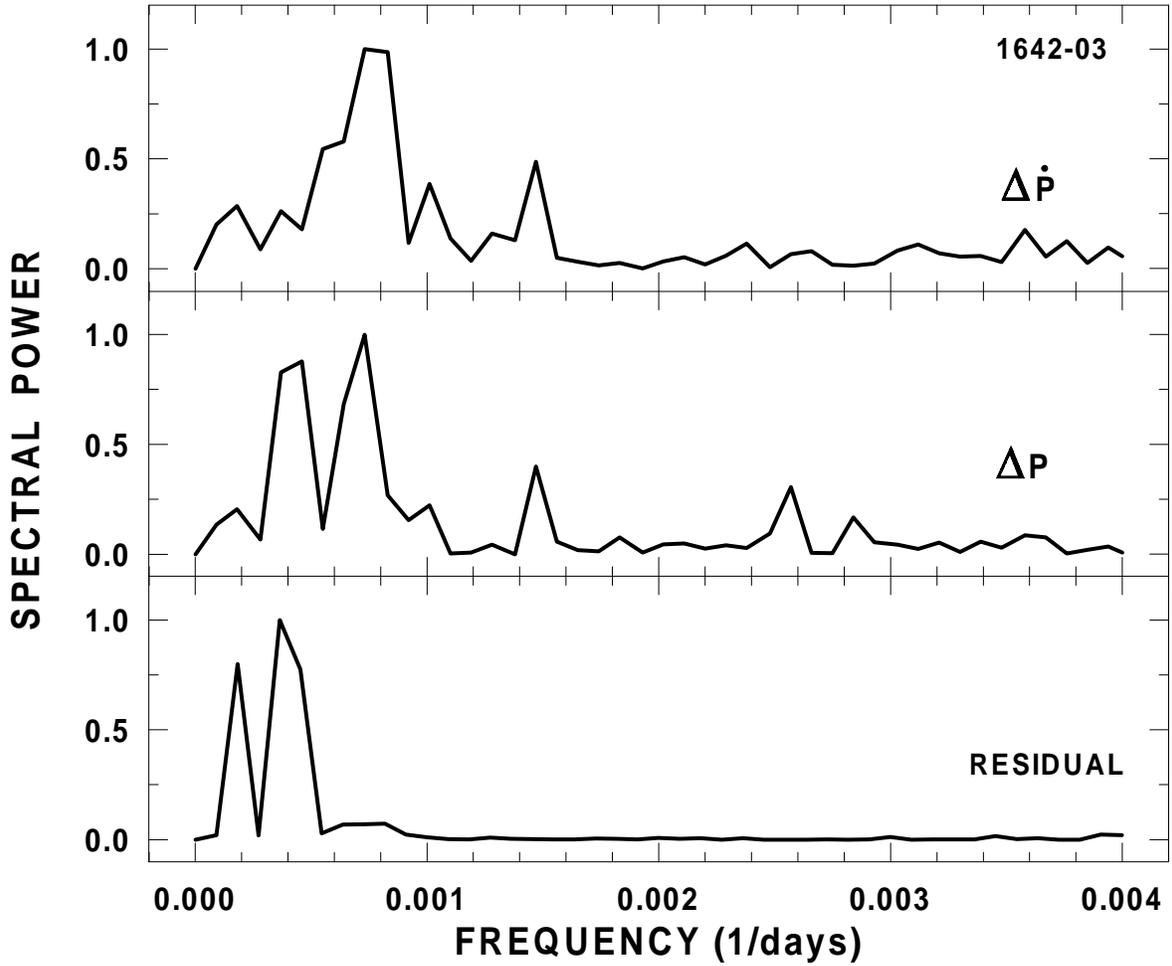}
   \caption{The power spectrum of the period derivative
         residuals, the period residuals and the timing residuals
         obtained by using the Fourier transforms tecnique.
         Spectral power of the residuals has arbitrary normalization.
         The spectra exhibit wide spectral features at multiple
         frequencies of approximately 0.0004 and 0.0008 $day^{-1}$,
         corresponding to 2500 and 1250  days respectively.
           \label{spectr}}
   \end{figure}

In order to find periodicities contained in these time sequences,
the power spectrum was computed using the Fourier transform.
The results are given in Figure~\ref{spectr}.
There are two wide spectral features at around 0.0004 and
0.0008 $day^{-1}$, which are visible in the spectrum of all the three
sequences. They have an identical amplitude in the spectrum of
$\Delta{P}$, whereas the spectrum of $\Delta{\dot{P}}$ exhibits
more distinctly the second feature at around 0.0008 $day^{-1}$ and
the residual spectrum shows more clearly the first feature at
0.0004 $day^{-1}$. These spectral frequencies are multiple
and correspond to the periods of 2500 and 1250 days respectively.
The features are wide because the timing residuals seen in
Figure~\ref{phres} are clearly not a strict periodic function of time.
The spectral feature at around 0.00018 $day^{-1}$ in the residual plot
could be due to the envelope of the timing residual curve.
The spectra of residuals $\Delta{\dot{P}}$ and $\Delta{P}$ also exhibit
the feature at around 0.0015 $day^{-1}$ corresponding to the
period of 667 days.

\placefigure{spectr}

\subsection{Analysis of multifrequency observations of the pulsar}
\begin{figure}
   \plotone{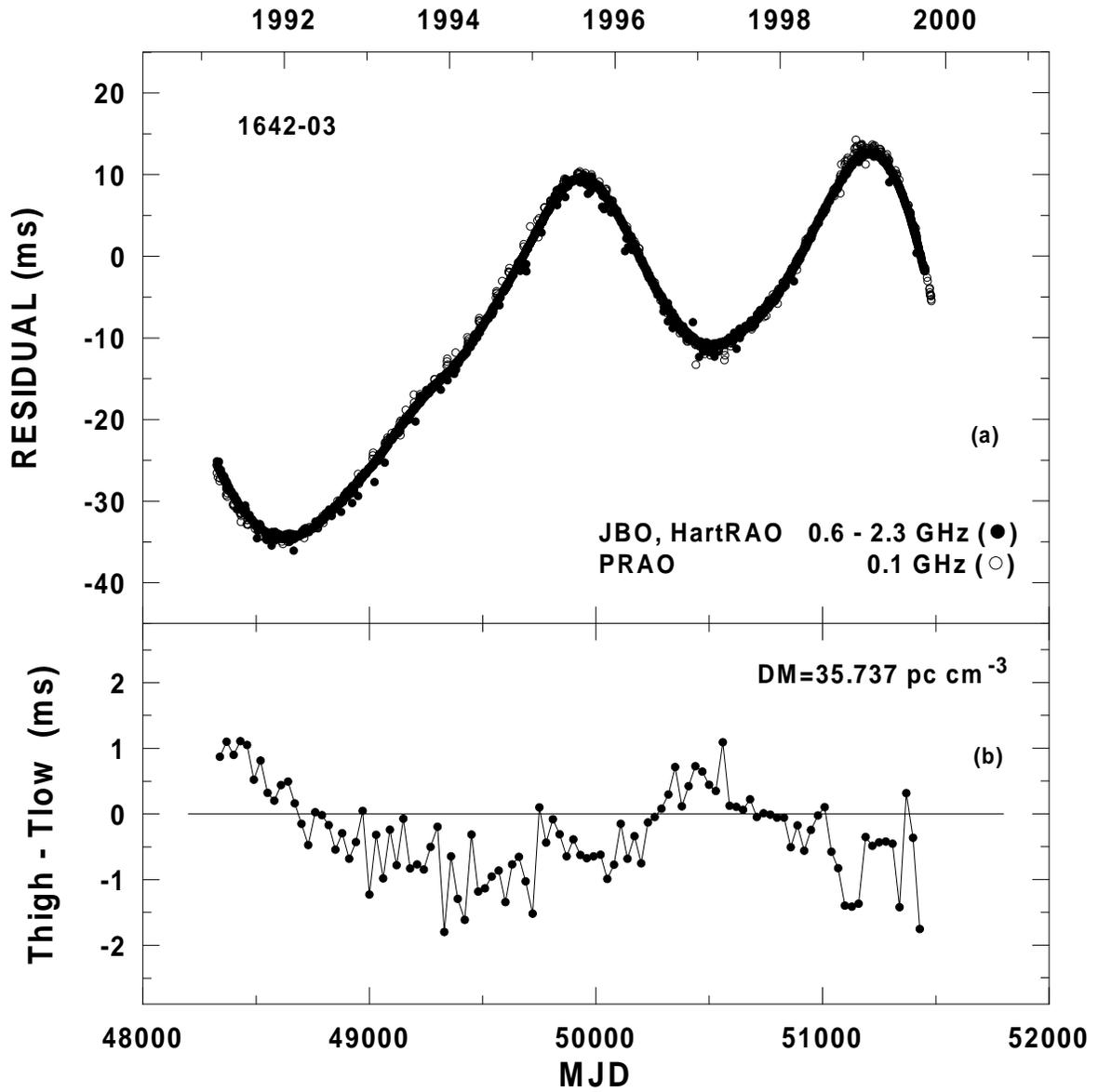}
   \caption{Top: the timing residuals of PSR B1642$-$03
           between 1991 March and 1999 October.  Bottom: the
           time-averaged differences in timing residuals obtained at
           high frequencies in the range 0.6 - 2.3 GHz and at the low
           frequency of 0.1 GHz.
           \label{multif}}
   \end{figure}

Since 1991, the pulsar rotation has been monitored by JBO, HartRAO
and PRAO and the timing residuals of PSR B1642$-$03 for these
quasi-simultaneous observations in the frequency range 0.1 - 2.3 GHz
are shown in Figure~\ref{multif}. The common span covers almost
two cycles of the oscillatory timing residuals.
Although the high- and low-frequency residual curves are similar
in general form, at some phases of the cycle there is an
appreciable time offset between the two different residual curves.
The time offset is calculated by subtracting the low-frequency
timing residuals from the high-frequency timing residuals averaged
over about 40 days. This time offset is shown in
Figure~\ref{multif}b and has a weak quasi-sinusoidal variation
with a maximum amplitude of $\sim$1 ms. This quasi-sine wave is, more or
less, out of phase with the residual curve shown in
Figure~\ref{multif}a. At the maximum of the timing residual curve,
the 0.1-GHz pulses arrive $\sim$1 ms later than the pulses at high
frequencies, and at the curve minimum, the 0.1-GHz pulses arrive
earlier than high-frequency pulses.

The 1-ms amplitude of the quasi-sinusoidal wave of the 0.1-GHz
residuals may be interpreted in two ways.  Firstly, it may arise
from a change in dispersion measure of $\sim2.5\times10^{-3}\ pc\
cm^{-3}$.  Alternatively, it may be interpreted as a result of
changes in the time alignment of the low-frequency and
high-frequency profiles during a cycle of the timing residuals. It
is very plausible that the beam shapes are different at low- and
high-frequencies so that the timing residuals could have different
shapes at the two frequencies.

\placefigure{multif}

\section{Discussion}
The pulsar B1642$-$03 is located above the galactic plane with
$b\sim26\arcdeg$ and has a large uncertainty in the distance
from the Sun. The pulsar catalog (\cite{tay93}) quotes the
dispersion measure-derived distance of 2900 pc with an uncertainty
of $\sim50\%$. On the other hand, Prentice and ter Haar
(1969)\nocite{pre69} found for PSR B1642-03 a distance of about
160 pc taking into account the presence of the HII region along the
line-of-sight to this pulsar and its influence on the dispersion
measure. Despite a large uncertainty in the distance, a small
proper motion suggests that this pulsar is a low-velocity
object. The observed value of $\mu=0\farcs002\pm0\farcs007\
yr^{-1}$ implies a transverse pulsar velocity equal to 2 $km\ s^{-1}$
and 30 $km\ s^{-1}$ for an assumed distances of 160 pc and 2900 pc
respectively. This low-velocity pulsar has a galactic z-distance of
70 pc and is inside the thin disk in which extreme Population I objects
are born.

The most likely explanation for the cyclical timing residuals for
the pulsar B1642$-$03 lies in the precession of a neutron star.
The spin axis of an isolated pulsar can precess, if a neutron star
has a non-spherical shape and its spin axis is not aligned with
the symmetry axis. Free precession will cause a cyclical change in
the angle $\alpha$ between the spin axis and the magnetic moment.
The result will be cyclical variations in the spin-down torque
acting on the pulsar and cyclical changes in the observed pulse
profile because the observer will view the pulsar beam from
different angles over the precession cycle (\cite{pin72};
\cite{shh77}; \cite{nel90}; \cite{cor93}). Suggestions that the
correlated timing behaviour and variations in the pulse shape
might be related with the precessional effects were discussed
earlier for some pulsars (\cite{cor93}; \cite{sul94};
\cite{ale97}). A unique result was recently obtained for the
pulsar PSR B1828$-$11 by Stairs, Lyne, \& Shemar
(2000)\nocite{sta00}. They reported the discovery of long-term,
highly periodic, correlated variations in both the pulse shape and
rate of slow-down of the pulsar PSR B1828$-$11. The authors explained
these periodic changes by precession of a neutron star spin
axis caused by an asymmetry in the shape of the pulsar.

In the case of PSR B1642$-$03, we observe long-term variations
only in the timing residuals and do not see any profile shape changes
in the range 0.1 - 1.6 GHz. Probably, the pulse shape
variations were not detected due to cyclical character of the changes
and their small magnitude. It should be noted that
an analysis of the JPL data at 2.3 GHz  done earlier
by Blaskiewicz (1991)\nocite{bla91} showed that the pulse
shape of the pulsar has slight periodic changes in the brightness of
the leading component with a period of about 1000 days. So, the
pulse shape changes do exist and their timescale agree well
with our result. As was mentioned above, the spectral analysis
of the period derivative residuals for PSR B1642$-$03 indicates
the presence of the periodicity of about 1250 days.
Therefore, the long-term
variations in the pulsar rotation can be explained in terms of free
precession model as was suggested earlier by Cordes (1993)\nocite{cor93}.
The measured 1-ms quasi-sinusoidal time offset between the high- and
low-frequency timing data over the cycle may also testify that it is
related to variations in the rotation of the pulsar.

In total, the spectra of the residuals for PSR B1642$-$03  exhibit
a few wide spectral features at multiple frequencies at around
0.0002, 0.0004 and 0.0008 $day^{-1}$ that correspond to the
periodicities of approximately 5000, 2500 and 1250 days
respectively. Note that these are some preferred frequencies as
the strict periodic behaviour of PSR B1642$-$03 is absent.
Nevertheless, the multiple character makes this result similar to
the result obtained for the pulsar PSR B1828$-$11 (\cite{sta00}),
the spectra of which for the residuals and the pulse shape
parameter show harmonically related sinusoids with periods of
approximately 1000, 500 and 250 days. The presence of the multiple
frequencies in the spectra of the residuals may testify by analogy
with PSR B1828$-$11 that it is related with precession of a
neutron star spin axis. The observed variations for PSR B1642$-$03
in the period derivative $\dot{P}$ make up 1.7$\%$. We may
estimate corresponding changes in the angle $\alpha$, using the
vacuum dipole model, in which the slow-down rate $\dot{P} \propto
\sin^{2}\alpha$. Suggesting that $\alpha\sim60\arcdeg$, the change
of similar magnitude in $\sin^{2}\alpha$ will give a variation in
the magnetic inclination angle $\alpha$ of 0.8 degrees. This value
is very similar to that obtained by Cordes (1993)\nocite{cor93}.

Observations of the long-term cyclical variations in the pulsar
spin for PSR B1642$-$03 may provide a valuable tool for
investigating the problems of the neutron star structure. The
occurrence of free precession in neutron stars is highly
problematic because of the superfluid interior (\cite{shh77};
\cite{shh86}; \cite{sed99}). Glitches, which have observed in
several dozen mostly young pulsars, provide strong evidence that
superfluid neutrons exist in neutron stars and these are expected
to damp out any free precession on timescales of several hundred
precession periods. However the clear correlated changes in pulse
shape and slowdown rate in B1828$-$11 and the observations
presented here indicate that precession is indeed occurring. A
model will have to be devised which allows for precession to
survive, even in the presence of some superfluid vortices.

\acknowledgments
TVSH and JOU are grateful to the Director of HartRAO G.D. Nicolson
for making the HartRAO pulsar data available to them, and to C.S.
Flanagan who set up the HartRAO pulsar programme and acquired most
of the data. TVSH thanks Yu.P. Shitov for valuable comments and
the staff of PRAO for help with the observations. This work was
supported by grant INTAS 96-0154. JOU acknowledges the hospitality
and support of HartRAO. IAU Commission 38 grant enabled him to
visit South Africa. The authors are grateful to the referee for
helpful comments and suggestions.

\clearpage

\begin{deluxetable}{ll}
\footnotesize
\tablewidth{33pc}
\tablecaption{Observed Parameters of PSR B1642-03
    \label{tbl-1}}
\tablehead{ \colhead{Parameter} & \colhead{Value\tablenotemark{a}} }

\startdata
Period, $P\ (s)$                          & 0.387688759475(4)    \nl
Period derivative, $\dot{P}\ (10^{-15}\ s\ s^{-1})$    & 1.780527(10)  \nl
Epoch of period (MJD)                     & 40414.1297           \nl
Right ascension, $\alpha$ (J2000) &
                        $16^{h}\ 45^{m}\ 02^{s}.045(3)$          \nl
Declination, $\delta$ (J2000)     &
         ${-03}{\arcdeg}\ 17{\arcmin}\ {58}{\arcsec}.35(10)$     \nl
$\mu_{\alpha}\ (mas\ yr^{-1})$            & 1(2)                 \nl
$\mu_{\delta}\ (mas\ yr^{-1})$            & 2(10)                \nl
Dispersion measure, $DM\ (pc\ cm^{-3})$   & 35.737(3)            \nl
\enddata

\tablenotetext{a}{Quoted errors are twice the formal standard errors
and refer to the least-significant digit.}
\end{deluxetable}


\newpage
\clearpage
%
\newpage
\clearpage

\newpage
\clearpage
\newpage
\clearpage
\newpage
\clearpage
\end{document}